\documentclass[nofootinbib,preprint,showkeys]{revtex4-1}
\usepackage{amsmath,amssymb}
\usepackage{graphicx}
\usepackage{slashed}

\newcommand{\be}{\begin{equation}}
\newcommand{\ee}{\end{equation}}

\begin{document}

\title{On the Observables of Renormalizable Interactions}
\author{Kang-Sin Choi}
\email{kangsin@ewha.ac.kr}
\affiliation{Scranton Honors Program, Ewha Womans University, Seoul 03760, Korea}
\affiliation{Institute of Mathematical Sciences, Ewha Womans University, Seoul 03760, Korea}
\begin{abstract}
We reconsider the renormalization of scalar mass and point out that the quantum correction to the physical observable, as opposed to the bare parameter, of a renormalizable operator, is technically insensitive to ultraviolet physics and independent of the regularization scheme. It is expressed as the difference in the same quantities at different energy scales, maintaining the same asymptotics. Thus, any sensible regularization cancels out the divergences, including the quadratic ones, and yields the same finite corrections.
To this end, we first show that the vacuum polarization of quantum electrodynamics is independent of the regularization scheme and a gauge-dependent quadratic divergence is canceled in the observable. We then calculate the quantum correction to the Higgs mass squared by the top-quark loop. It is again finite and regularization-scheme independent. For large external momentum, the correction of the pole mass-squared is dominated by power running, resulting in an order of 1 percent correction. In particular, the effect of heavy fields on the scalar mass correction is suppressed.
\end{abstract}
\maketitle

To address the problem of infinity arising from the radiative corrections in quantum field theory, we need to focus on what we can observe.
A bare parameter in the Lagrangian is not an observable, because interactions modify it, without which we cannot probe it. We always observe a {\em combination} of a bare parameter and its corrections, which is inevitably a function of the energy scale. In quantum field theory, such correction gives rise to divergence since we include virtual particles with the energy-momentum up to infinity to preserve the necessary symmetry for calculation. However, the divergent correction from virtual particles to an unobservable bare parameter gives us a wrong impression; we have to convert it into {\em observable} one.

An observable can be given as a {\em difference} of the corrected quantities between different energy scales, which does not include the bare parameter. In this letter, we show that the divergent parts are {\em canceled} in the observable, regardless of the regularization scheme that quantifies the divergence, yielding only finite observables.
We take two examples: First, we consider the electric charge corrected by the self-energy of the photon that contains illusory quadratic divergence. Then, we calculate the correction to a scalar mass, for which it is believed that there is a fundamental quadratic divergence.

Consider a photon described by the propagator in the Feynman gauge,
\be
 \Delta^{0}_{\mu\nu}(q^2) = \frac{-i g_{\mu\nu}}{q^2 + i \epsilon} .
\ee
In what follows, we omit the Feynman prescription $i\epsilon$. 
Letting one-particle-irreducible self-energy $i{\Pi^{\mu}}_\nu (q^2)$, we can only detect the sum of its powers
\be \label{Pisum} \begin{split}
 {\Delta^{\prime \mu}}_\nu (q^2)&= \frac{ -i\delta^\mu_\nu}{q^2}  +\frac{ -i \delta^\mu_\rho}{q^2}  [i{{\Pi^\rho}_\sigma}(q^2)] \frac{ -i \delta^\sigma_\nu}{q^2}  + \dots\\
  & = -i  [q^2 {\delta_\mu^\nu}- {{\Pi^\nu}_\mu}(q^2) ]^{-1}.
\end{split}
\ee
We do not assume the usual pole at $q^2=0$; we consider a nontriviality ${{\Pi^\mu}_\nu}(0)\ne 0$ that arises from quadratic divergence due to improper regularization.

In ${\Pi^{\mu}}_{\nu}(q^2)$, the term proportional to $q^\mu q_\nu$ depends on the choice of the gauge, so we may neglect them for the moment and consider the component proportional to ${g^{\mu}}_\nu$
\be \label{proportionality}
 {\Pi^{\mu}}_\nu(q^2) \equiv{\Pi^{\mu}}_\nu(0)+ \delta^{\mu}_\nu q^2 \Pi(q^2)+\text{(gauge terms)}.
\ee
Then, the self-energy is truly a function of the Lorentz-invariant quantity $q^2$ and essentially the index structure becomes that of identity. 
For this, we need the Ward--Takahashi identity \cite{Ward:1950xp} that we cannot formally use at this moment, but we justify it later in Eq. (\ref{WeakWard}). We are going to care about the divergent term not obeying it.

First, we question whether it is possible to observe the bare parameter.
To understand this, let us take a reference point, where the propagator becomes
\be
 q^2 \to 0: {{\Delta^{\prime }}^\mu}_\nu (q^2) \to -i \left[q^2  {\delta_\mu^\nu}-{{\Pi^\nu}_\mu}(0)-q^2\frac{d{{\Pi^\nu}_\mu}}{d(q^2)}(0)\right]^{-1}.
\ee
In the case we have a pole at $q^2=0$, the corresponding residue makes undesirable normalization. The field renormalization can absorb it
\be \label{renormalization}
 {Z_{3\mu}}^\nu \equiv \left[{\delta_\nu^\mu}- \frac{d{{\Pi^\mu}_\nu}}{d(q^2)}(0) \right]^{-1}, \quad
 A_\mu (x) \to {\sqrt{Z_3}_\mu}^\nu A_\nu(x),
\ee
for which we understand that the square root of the matrix has positive entries.
This is possible because we cannot observe $A_\mu(x)$ directly. Here, this process is meaningful when there is a pole at $q^2=0$, or ${{\Pi^\mu}_\nu}(0)=0$, but the renormalization \eqref{renormalization} is always possible.

In terms of the renormalized gauge field, the propagator becomes, to ${\cal O}(q^4)$,
\be
 \begin{split}
{\Delta^\mu}_\nu (q^2)= -i &[q^2 {\delta_\mu^\sigma}- {{\Pi^\sigma}_\mu}(q^2) ]^{-1} \left[ {\delta_\nu^\sigma}- \frac{d{{\Pi^\sigma}_\nu}}{d(q^2)}(0) \right] .
\end{split}
\ee

With the bare charge $e_0$, the electric charge can be seen by an interaction involving $e_0^2 {\Delta^\mu}_\nu(q^2)$. An interacting field sees the effective charge $e(q^2)$ at energy $q^2$ without knowing what is happening inside and it recognizes the propagator ${\Delta^{0\mu}}_\nu(q^2)$ as 
\be
 e^2(q^2){\Delta^{0\mu}}_\nu (q^2)= e_0^2{\Delta^\mu}_\nu (q^2),
\ee
(dressed by a vertex function and external fermions) or,
\be
 e^2(q^2) {\delta_\nu^\mu}\equiv e_0^2  \left[{\delta_\nu^\mu} + \frac{{{\Pi^\mu}_\nu}(q^2)}{q^2} 
 - \frac{d{{\Pi^\mu}_\nu}}{d(q^2)}(0) \right].
\ee
Taking $q^2\to 0$ and performing Taylor expansion, we have
\be
 q\to0:  e^2(q^2){\delta_\nu^\mu} = e_0^2 \left[{\delta_\nu^\mu} +\frac{{\Pi^\mu}_\nu(0)}{q^2}  \right].
\ee
This enables us to remove the dependence on the bare coupling $e_0$.
Finally, the observed charge in terms of the reference charge $e^2(0)$ is
\be \label{combination}
\begin{split}
e^2(q^2) {\delta_\nu^\mu}&= e^2(0)  \left[{\delta_\nu^\mu} + \frac{{{\Pi^\mu}_\nu}(q^2)-{{\Pi^\mu}_\nu}(0)}{q^2}-\frac{d{{\Pi^\mu}_\nu}}{d(q^2)}(0) \right]\\
& \equiv e^2(0) \left[{\delta_\nu^\mu} + \frac{\delta{{\Pi^\mu}_\nu}(q^2)}{q^2} \right] .
\end{split}
\ee
This means that we are only able to see {\em the difference} of the self-energy\footnote{A technicality: To maintain the covariant form, the differentiation should be modified from $q^2\frac{d\Pi^{\mu \nu}}{d(q^2)}(0)$ to $q^2\frac{d\Pi^{\mu\nu}}{d(q^2)}|_{q^\mu q^\nu \text{fixed}}(0)+ q^\mu q^\nu \frac{d\Pi^{\mu \nu}}{d(q^\mu q^\nu)}|_{q^2 \text{fixed}}(0)$.}
\be \label{ModSelfE}
 \delta {{\Pi^\mu}_\nu}(q^2) = {{\Pi^\mu}_\nu}(q^2) -{{\Pi^\mu}_\nu}(0)- q^2\frac{d{\Pi^\mu}_\nu}{d(q^2)}(0)
\ee
from a certain reference point, as the charge correction, through a scattering process. 

\begin{widetext}
For concreteness, we calculate the one-loop correction to the self-energy due to a fermion of mass $m$ \cite{Peskin:1995ev},
\be \label{Pi2}
  i  \Pi^{\mu \nu}_2(q^2)=  -4e^2  \int_0^1 dx \int \frac{d^4 \ell}{(2\pi)^4} \frac{1}{\left( \ell^2 -\Delta \right)^2}\left[- \frac12 g^{\mu \nu} \ell^2 -2x(1-x)q^\mu q^\nu + g^{\mu \nu} (m^2+x(1-x)q^2)\right],
\ee
in the Minkowski momentum $\ell$, where $\Delta \equiv m^2 - x(1-x) q^2$. We used the reference charge $e^2 \equiv e^2(0)$ and the reference mass $m$.
Instead of dimensional regularization \cite{tHooft:1972tcz} that is known to give no quadratic divergence, let us take a simple cutoff scheme that respects little symmetry, giving
\begin{align}
   \int \frac{d^4 \ell}{(2\pi)^4} \frac{\ell^2}{\left( \ell^2 -\Delta \right)^2}
   &=\frac{i}{16\pi^2} \left(-  \Lambda^2 + 2 \Delta \log \frac{\Lambda^2}{ \Delta} - \Delta\right),  \label{cutoff1} \\
 \int \frac{d^4 \ell}{(2\pi)^4} \frac{1}{\left( \ell^2 -\Delta \right)^2} 
 &= \frac{i}{16\pi^2}\left( \log \frac{\Lambda^2}{\Delta} - 1 \right) \label{cutoff2},
 \end{align}
with the (Euclidianized) cutoff at $-\ell^2=\Lambda^2$.

Usually, this regularization scheme cannot respect the gauge symmetry, leading to spurious quadratic divergence. However, we can only see the difference as in \eqref{ModSelfE}. 
Thus we have\footnote{See Footnote 1.}
\be \label{deltaPi2} \begin{split}
 \delta \Pi^{\mu \nu}_2(q^2) 
 &  =- \frac{e^2}{4\pi^2} \int_0^1 dx \left[ \frac12 g^{\mu \nu}(\Lambda^2 - \Delta) - 2x(1-x) (q^\mu q^\nu - g^{\mu \nu} q^2)\left( \log \frac{\Lambda^2}{ \Delta} -1 \right) \right. \\
   &\ -  \left. \frac12 g^{\mu \nu} (\Lambda^2 - m^2)+ 2x (1-x) (q^\mu q^\nu - g^{\mu \nu} q^2)\left( \log\frac{\Lambda^2}{ m^2}  -1 \right) - \frac{1}{2} g^{\mu \nu}x (1-x) q^2  \right] \\
   & =-\frac{e^2}{2\pi^2} (g^{\mu \nu} q^2 - q^\mu q^\nu) \int_0^1 dx  x(1-x) \log \frac{m^2}{m^2-x(1-x)q^2}.
\end{split}
\ee
\end{widetext}
This is the same result as that obtained by the dimensional regularization \cite{Peskin:1995ev}.

In $\Pi_2^{\mu \nu}(q^2)$, a constant term $g^{\mu \nu} \Lambda^2/2$ is generated, which we tend to interpret as the infinite photon mass. Of course, this is scheme-dependent and unphysical because we know that the gauge symmetry protects the photon from being massive. Remarkably, that constant term {\em is absent in the observable} difference $\delta \Pi_2^{\mu \nu}(q^2)$ in \eqref{deltaPi2}; quantum field theory still takes good care of it and does give the correct observable in the end. We have a weak form of the Ward--Takahashi identity on the observable
\be \label{WeakWard}
q_\mu \delta \Pi_2^{\mu \nu}(q^2) =q_\nu \delta \Pi_2^{\mu \nu}(q^2)=0.
\ee
This justifies the claim of unobservability of $q^\mu q^\nu$ dependent terms around Eq. \eqref{proportionality}. 

It is interesting to compare how the dimensional regularization deals with the divergence. It gives, to ${\cal O}(\epsilon)$,
\begin{align}
 \int \frac{d^d \ell}{(2\pi)^d} &\frac{\ell^2}{\left( \ell^2 -\Delta \right)^2}  =\frac{i}{16\pi^2} \Delta \left(\frac{4}{\epsilon} - 2  \log \Delta + c  \right), \label{DimRegQuad}\\
 \int \frac{d^d \ell}{(2\pi)^d} &\frac{1}{\left( \ell^2 -\Delta \right)^2} = \frac{i}{16\pi^2}\left(\frac{2}{\epsilon} -\log \Delta + c' \right),
\end{align}
where $4-d\equiv \epsilon$. There are scheme-dependent constants $c$ and $c'$, however, they are canceled in the observable.
Identifying $\log \Lambda = 1/\epsilon$ up to a constant is identical to the above cutoff scheme, except that the dimensional regularization neglects the quadratic divergence or makes it logarithmic. It has a merit in that we can see the gauge invariant structure already in $\Pi^{\mu \nu}_2(q^2)$, but this makes {\em no difference in the observable} \eqref{ModSelfE}. 

In general, the divergences are canceled in the following way \cite{BPHZ}.
\begin{enumerate}
\item
If a divergence does not depend on $q^2$, the difference ${{\Pi^\mu}_\nu}(q^2)-{{\Pi^\mu}_\nu}(0)$ cancels it. This is the case for the quadratic divergence.
\item
If a divergence depends on $q^2$, then we should be able to separate it from ${{\Pi^\mu}_\nu}(q^2)-{{\Pi^\mu}_\nu}(0)$ in the limit $q^2 \to 0$, that is, the divergence is contained in the differentiation $q^2 \frac{d{\Pi^\mu}_\nu}{d(q^2)}(0)$.  The field renormalization cancels it. This is the case for the logarithmic divergence.
\end{enumerate}
Dimensional analysis shows that this should hold true independent of the order of perturbation (If we use momentum-dependent renormalized parameters as ones used in this letter, there is no corresponding subdiagram divergence \cite{Weinberg:1959nj}). This generic feature applies to any sensible regularization preserving good enough symmetry. In particular, we have not relied on the gauge symmetry to remove the quadratic divergence. 

The modern understanding of the cutoff $\Lambda$ is the upper limit of the energy scale to which the theory at hand is valid \cite{Wilson:1971bg}. The ultraviolet asymptotics, or the $\ell^2$ integral around $\Lambda^2$, of the functions ${{\Pi^\mu}_\nu}(q^2)$ and ${{\Pi^\mu}_\nu}(0)$ are the same in the limit $q^2\to 0$, or the accuracy is ${\cal O}(q^2/\Lambda^2)$.


Now, we turn to the scalar mass. We have a Feynman propagator for a scalar field
\be \label{FeynProp}
 D_F^0 = \frac{i}{p^2- m_0^2 }
\ee
encoding the information on the bare mass $m_0$. However, this is only revealed by interactions with other particles, which correct the bare mass
\be
 D_F^{\prime }= \frac{i}{p^2- m_0^2-\tilde \Sigma(p^2)},
\ee
having the same geometric sum structure by the one-particle-irreducible self-energy $\tilde \Sigma(p^2)$ as in (\ref{Pisum}).
The mass changes and becomes dependent on the external momentum $p^2$. We take the pole mass $m_h$ as a reference 
\be \label{polemass}
 [ p^2 - m_0^2 - \tilde \Sigma(p^2)]_{p^2=m_h^2} =0.
\ee 
This is a usual renormalization condition and corresponds to parameter fitting in the effective field theory.
Also, the residue is changed, which comes from the next leading order expansion from $\tilde \Sigma(p^2)$ from a reference mass
\be
 p^2 \to m_h^2:  \frac{i}{p^2- m_0^2-\tilde \Sigma(p^2)} \to \frac{iZ}{p^2-m_h^2}+ {\cal O}((p^2-m_h^2)^2),
\ee
where
\be
 Z^{-1} = 1- \frac{ d \tilde \Sigma}{d p^2}(m_h^2).
\ee
To recover the desired one, we renormalize the field $\phi(x) \to \sqrt{Z} \phi(x)$ as before. 

To have the desired pole, we need to define the momentum-dependent mass accurately to  order of $p^2-m_h^2$ (\cite{Coleman:2018mew}, Eq. (15.56))
\be \label{renormalizaedmass}
 m^2(p^2) = m_0^2 + \tilde \Sigma(p^2) - (p^2 -m_h^2) \frac{ d \tilde \Sigma}{d p^2} (m_h^2)
\ee
Since the observable, the finite pole mass (\ref{polemass}),
$$
 m_h^2 = m_0^2 +\tilde \Sigma(m_h^2),
$$
shows that neither the bare mass $m_0^2$ nor the correction $ \tilde \Sigma(p^2)$ is finite; therefore, we do not attempt to explain the {\em absolute smallness of $ \tilde \Sigma(p^2)$} compared to the Planck scale, employing symmetry (the ``big'' hierarchy problem). Rather, we calculate the quantum {\em correction} from $m_h^2$, which gives us a prediction on observable mass and see whether it can consistently {\em remain small} (the technical hierarchy problem).

In terms of the pole mass (\ref{polemass}), we have the expression for the observable mass \eqref{renormalizaedmass}, 
\be \label{SlidingMass}
\begin{split}
  m^2(p^2) &= m_h^2 + \tilde \Sigma(p^2) - \tilde \Sigma(m_h^2) - (p^2 -m_h^2) \frac{ d \tilde \Sigma}{d p^2} (m_h^2)\\
  & \equiv m_h^2+ \delta m_h^2(p^2)
\end{split}
\ee
We can only measure the energy-dependent mass $m^2(p^2)$ from the reference $m_h^2$. 

It has the same structure as the photon self-energy (\ref{ModSelfE}), canceling quadratic and logarithmic divergences, which is scheme-independent.
This is similar to the on-shell renormalization; however here, we do not use counterterms and the mass slides whenever we include a correction. 

The self-energy $\tilde \Sigma(p^2)$ contains the contributions from various fields at various orders of perturbations.
The largest contribution comes from the one-loop amplitude of the top quark (\cite{Peskin:1995ev}, Eq. (10.33))
\be
 - i \tilde \Sigma^t_2(p^2) =-2  y_t^2 N_c \int_0^1 dx \int\frac{d^4 \ell}{(2\pi)^4} \frac{\ell^2-x(1-x)p^2+m_t^2}{(\ell^2+ x(1-x)p^2-m_t^2)^2},
\ee
where $y_t$ is the diagonalized top quark Yukawa coupling, $N_c=3$ is the number of colors \cite{Feng:2013pwa} and $m_t$ is the top quark mass. This quantity is divergent, but we can observe only {\em the difference} in (\ref{SlidingMass}). 
Note that the contribution from the top-quark loop $\tilde \Sigma^t_2(p^2)$ can account for the matching (\ref{polemass}) to accuracy ${\cal O}(y_t^2)$. 

Using {\em any} of the above regularizations, we obtain the mass correction
\be \label{masscorrection}
 \delta m_h^2(p^2) = -\frac{3 y_t^2}{16\pi^2} \left[p^2-m_h^2+ 6 \int_0^1 dx (m_t^2-x(1-x) p^2) \log \frac{m_t^2-x(1-x)p^2} {m_t^2-x(1-x)m_h^2}\right] + {\cal O}(y_t^3),
\ee
which is finite; it is free of the divergent quantity $\Lambda$ and expressed in terms of the finite parameters $m_t^2$ and $m_h^2$. 

Note that the scalar mass shows power running in $p^2$, which we regard as the sliding scale. Because the chiral symmetry does not protect it, a significant correction is possible. The mass correction is real for $p^2<4m_t^2$, above which the Higgs decays to a top pair. The momentum is not extended to an arbitrarily high scale and the top quark is the ``new physics.'' 
 
\begin{figure}[t]
\includegraphics[scale=0.7]{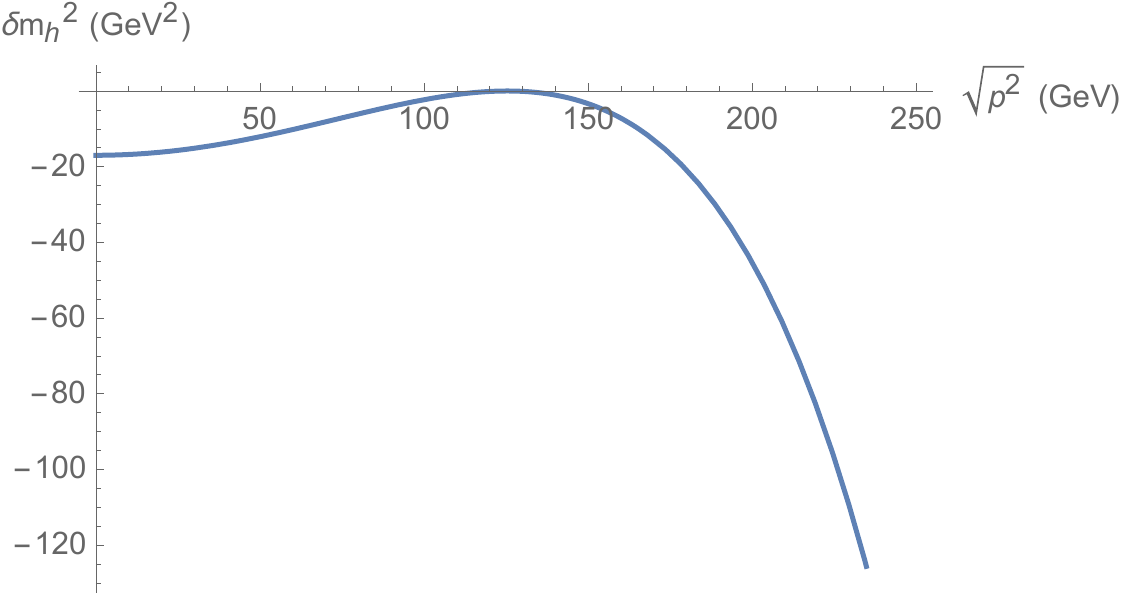}
\caption{\label{fig:mhcorr}The correction to the Higgs mass squared from the top quark loop, as a function of the relativistic energy $\sqrt{p^2}$. We use $m_h = 125.4, m_t = 173$, all in GeV, and $y_t=1$. For large $p^2$, the power running becomes dominant and the correction is relatively large. For $\sqrt{p^2}=250\, \rm GeV$, the mass correction is $\delta m_h^2 = - 181.78 \, \rm GeV^2$ or $-1.156\, \%$.}
\end{figure}
We plot the correction to the Higgs mass squared for typical parameters in Fig. \ref{fig:mhcorr}. 

Now, we clarify the technical hierarchy problem. First, the cutoff dependence is small. The use of a bare parameter gives us an illusory dependence on $\Lambda$. 
A quadratic divergence in $\tilde \Sigma(p^2)$ in the cutoff scheme is canceled. The loop correction should not be scheme-dependent. There is no miraculous cancellation of ultraviolet scale dependence between the bare parameter and the correcting self-energy as in (\ref{renormalizaedmass}). An equivalent description is in (\ref{SlidingMass}), where there is the cancellation of the same quantities at two different scales, to accuracy ${\cal O}((p^2-m_h^2)/\Lambda^2)$. The cutoff $\Lambda$ used in the regularization can also be regarded as the upper bound of the energy scale, e.g., $\sqrt{p^2}$ in (\ref{SlidingMass}), to which the description of the Higgs field in the Standard Model is valid \cite{Wilson:1971bg}. We can match new physics parameters, for instance, the mass of a heavy field, to the parameters at $\Lambda$. 
In the above Higgs mass correction (\ref{masscorrection}), the dependence of the $\Lambda$ is canceled to accuracy ${\cal O}((p^2-m_h^2)/\Lambda^2)$. The mass correction is not sensitive to high energy cutoff to this accuracy.

Secondly, the contribution from heavy fields is suppressed. The mass correction formula \eqref{masscorrection} shows exemplar features of the decoupling \cite{Appelquist:1974tg}. 
In the Standard Model, the top quark mass and the Yukawa coupling are proportionally related $m_t \propto y_t$. However, we may formally regard them as independent and treat the top quark as a heavy fermion at high, e.g. the Grand Unification, scale. Then, the mass correction \eqref{masscorrection} is suppressed as
\be
 \delta m_h^2(p^2) \simeq -\frac{3 y_t^2(p^2-m_h^2 )^2}{16\pi^2} \left[ \frac{1}{10 m_t^2} + \frac{p^2+2m_h^2}{140 m_t^4}+\dots \right],
\ee
and this fermion {\em decouples} for $m_t\gg1$. The decoupling also occurs for an additional heavy scalar \cite{Choi:2023mma}. 

To conclude, the loop corrections to the scalar mass-squared from other fields are insensitive to the new physics parameter and are suppressed if they are very heavy. This supports the rationale for the smallness of $m_h^2$ against quantum corrections.
The essence of renormalization lies in the energy dependence of a physical parameter, which arises from a finite correction from finite parameters.

\begin{acknowledgments}
The author is grateful to Kiwoon Choi, Hyung Do Kim, Bumseok Kyae, and Hans-Peter Nilles for discussions. This work is partly supported by the National Research Foundation of Korea grant RS-2023-00277184.
\end{acknowledgments}

\end{document}